\documentclass[letterpaper,11pt]{article}
\usepackage{authblk}
\usepackage{amssymb}
\usepackage{tabularx}
\usepackage{subcaption}
\usepackage{multirow}
\usepackage{amsmath}
\usepackage{graphicx}
\usepackage[margin=1in,letterpaper]{geometry} 
\usepackage{cite}
\usepackage[final]{hyperref}
\usepackage[table]{xcolor}
\usepackage{color}
\usepackage{listings}
\usepackage[toc,page]{appendix}
\usepackage{docmute}

\makeatletter
\newcommand*{\rom}[1]{\expandafter\@slowromancap\romannumeral #1@}
\makeatother
\bibliography{journals,phd-references} 
\graphicspath{{images/}}
\DeclareGraphicsExtensions{.jpg}


\hypersetup{
	colorlinks=true,       % false: boxed links; true: colored links
	linkcolor=blue,        % color of internal links
	citecolor=blue,        % color of links to bibliography
	filecolor=magenta,     % color of file links
	urlcolor=blue         
}

\begin{document}

\title{Two-dimensional correlation function of binary black hole coalescences}
\author[1]{Marco Cavagli\`{a}\footnote{Corresponding author. Email: cavagliam@mst.edu}}
\author[2]{Ashini Modi}
\affil[1]{Institute of Multi-messenger Astrophysics and Cosmology, Missouri University of Science and Technology, Physics Building, 1315 N.~Pine St., Rolla, MO 65409, USA}
\affil[2]{Caddo Parish Magnet High School, 1601 Viking Dr., Shreveport, LA 71101, USA}
\date{\today}
\maketitle
\begin{abstract}
We compute the two-dimensional correlation functions of the binary black hole coalescence detections in LIGO-Virgo's first and second observation runs. The sky
distribution of binary black hole coalescence events is tested for correlations at different angular scales by comparing the observed correlation function to two
reference functions that are obtained from mock datasets of localization error regions uniformly distributed in the sky. No excess correlation at any angular scale
is found. The power-law slope of the correlation function is estimated to be $\gamma= 2.24\pm 0.33$ at the three-$\sigma$ confidence level, a value consistent with
the measured distribution of galaxies.
\end{abstract}

%%%%%%%%%%%%%%%%%%%%%%%%%%%%%%%%%%%%%%%%%%
\section{Introduction}\label{sec:Introduction}

On 14 September 2015, researchers from the Laser Interferometer Gravitational-wave Observatory (LIGO) \cite{TheLIGOScientific:2014jea} Scientific Collaboration
(LSC) and the European Virgo Collaboration \cite{Acernese:2015gua} made the first direct detection of Gravitational Waves (GWs) from a pair of coalescing black
holes \cite{Abbott:2016blz}. Less than two years after that first announcement, LIGO and Virgo observed GWs from the merger of two neutron stars
\cite{TheLIGOScientific:2017qsa}, an event that was rapidly followed by the Fermi Gamma-ray Space Telescope's detection of a gamma-ray flash, and eventually by
optical, infrared, radio, and X-ray observations by hundreds of telescopes around the world in what became the most observed event in the history of modern
astronomy \cite{GBM:2017lvd}. 

Currently, GW astronomy is a well-established scientific discipline. In the first two observing runs of Advanced LIGO and Virgo, O1 and O2, LSC and Virgo
collaboration researchers observed ten Binary Black Hole (BBH) coalescence detections and one binary neutron star coalescence detection. The~third observation run,
O3, brought us candidate detections on a weekly basis \cite{gracedb}, enabling a plethora of novel astrophysical and theoretical investigations. 

The next decade will see GW astronomy further expand its reach in frontier scientific research. Japan's KAGRA  detector \cite{Akutsu:2018axf} has joined the
international network of GW ground-based observatories. India has established the LIGO-India Scientific Collaboration (LISC) and finalized plans for the
construction of the LIGO-India detector \cite{LIGO-India}. The European space-based LISA  mission \cite{LISA}, slated to launch in 2034, will greatly improve
detection capabilities and localizations of astrophysical sources. The International Pulsar Timing Array project will detect ultra-low frequency GWs within ten
years~\cite{Hobbs:2009yy}. Optical, particle, and GW astronomy will together explore the Universe through complementary physical carriers.

The publication of LIGO-Virgo's first catalog of compact binary merger signals \cite{LIGOScientific:2018mvr} has shown that GW astrophysics is a powerful tool for
population and source property studies of compact objects, tests~of General Relativity, and large-scale cosmological measurements. However, many open questions
still remain. For example, tests of GR have returned a null result \cite{LIGOScientific:2019fpa}; the formation channels of black hole binaries
\cite{LIGOScientific:2018jsj} and the physics of EM-bright mergers \cite{LIGOScientific:2019eut} are still unclear, as well as the determination of the Hubble
constant from GW sirens \cite{Abbott:2019yzh}. Despite LIGO and Virgo running all-sky, unmodeled searches \cite{Abbott:2019prv,Abbott:2019heg}, no GW signal has
been detected that cannot be modeled as a compact binary coalescence. Other sources of multi-messenger signals such as isolated compact objects
\cite{Pisarski:2019vxw}, core-collapse supernovae \cite{Abbott:2019pxc}, and magnetars \cite{Abbott:2019dxx} have not been observed in the GW domain. 

The rapid growth in the number of BBH coalescence detections and the dramatic improvement in their sky localizations are turning GW astrophysics into a {precision}
observational science like large-scale structure astrophysics and early-Universe cosmology. One important physical concept in large-scale structure investigations
and observational cosmology is that of the {Correlation Function} (CF) \cite{White:1994}. The~(two-point) (auto-)CF describes the excess probability of finding
pairs of points at a given separation. In large-scale astrophysics and observational cosmology, the CF (or its homolog in the frequency space, the {power spectrum})
is commonly used to describe the spatial distribution of galaxies or the density fluctuations observed in the cosmic microwave background. The CF from galaxy
surveys, for example, allows astronomers to estimate the distance scales of galaxy clustering and gain information about the origin and evolution of the Universe's
large-scale structures.

The purpose of this short article is to introduce the concept of the CF for BBH coalescence events. We use the public BBH coalescence detections in LIGO-Virgo's O1
and O2 runs to compute the two-dimensional CF for the population of these objects. Ten detections with sky localizations ranging from 39 square degrees to 1666
square degrees are clearly not sufficient to draw any meaningful conclusion on the spatial distribution of BBH coalescences. However, this calculation shows that
the CF can be used to investigate the statistical properties of the population of these objects. We~illustrate the method by comparing the two-dimensional CF
obtained from the LIGO-Virgo O1-O2  BBH detections to a CF obtained by a random distribution of the same detections. The result shows that the two-dimensional
spatial distribution of the detections is consistent with an isotropic distribution, as~reported in Ref.~\cite{Stiskalek:2020wbj} by implementing a
pixelization-based method for the O1-O2 BBH detections. We~also confirm this conclusion by comparing the CF to a synthetic CF obtained by simulating a number of BBH
detections with sky localization error regions consistent with those of the LIGO-Virgo~sample.

\section{Two-Dimensional Correlation Function}\label{sec:sec}

In our analysis, we follow the customary definition for the two-dimensional (angular) CF of large-scale astrophysics \cite{White:1994}. The two-dimensional CF of a
population of objects describes the excess probability of finding two objects separated by the angular distance $\theta$ with respect to a uniform distribution. To
compute the CF of the BBH population, we treat the sky localization error regions of the BBH detections as probability density heat maps. Given the (normalized) sky
localization error region map of the $i^{\text{th}}$ BBH detection in the sample, $M_{i}(\chi,\varphi)$, where $\chi$ and $\varphi$ are the polar and azimuthal angles on the celestial sphere, respectively, we define the sky localization probability density map of the sample as:
\begin{equation}
M(\chi,\varphi) = \frac{1}{A(N)}\sum_{i=1}^N F_i(\chi,\varphi) M_i(\chi,\varphi)\,,
\label{eq:map}
\end{equation}
where $N$ is the number of BBH detections, $F_i$ are probability weights that depend on the GW detector network sensitivity, and $A(N)$ is a normalization factor. By expanding the sky localization map in
spherical harmonics,
\begin{equation}
M(\chi,\varphi) = \sum_{lm} a_{lm}Y_{lm}(\chi,\varphi)\,,
\label{eq:spherical_harmonics}
\end{equation}
the sky correlation function of the BBH sample can be defined as:
\begin{equation}
C(\theta) = \langle M(\hat n_1)\cdot M(\hat n_2)\rangle_{21}\,,
\label{eq:CF_defined}
\end{equation}
where the average is taken over the observed sky with angular separation $\theta$ held fixed. Using the addition theorem of spherical harmonics, the CF can be written as:
\begin{equation}
C(\theta) = \frac{1}{4\pi}\sum_la_l^2P_l(\cos\,\theta)\,,
\label{eq:CF}
\end{equation}
where $P_l(\cos\theta)$ denotes the Legendre polynomial of order $l$ and argument $\cos\theta$, and we have defined $a_l^2=\sum_m|a_{lm}|^2$. Note that the CF in
Equation~(\ref{eq:CF}) differs with the usual definition of the angular power spectrum that is used in Cosmic Microwave Background (CMB) cosmology, where $a_l^2 = (2l+1)C_l$. As~the map $M(\chi,\varphi)$ describes a probability density field, rather than the perturbation field of a physical quantity, in the following, we focus on the CF
instead of the power spectrum, which is the standard measure for fluctuation fields. 

The quantities $a_l^2$ are measured from the sky localization map $M(\chi,\varphi)$ and determine the two-dimensional angular distribution of the BBH sample. Comparison
of the CF to theoretical models involves modification of Equation~(\ref{eq:CF}) by multiplying the $a_l^2$ coefficients by a {window function} $W_l$ to take into account
experimental constraints in the observations. For example, the finite beam resolution of the detector introduces a high-$l$ cutoff that can be modeled with a window
function $W_l\propto\exp[-l(l+1)\sigma^2]$, where $\sigma$ is the detector resolution \cite{White:1994}. If the object population cannot be observed across the full
sky, a mask is required. In contrast with CMB observations, where the region of the sky along the galactic plane must be masked in CMB observations due to the
impossibility of measuring temperature fluctuations along the galactic plane, the full sky is transparent to GWs, and no mask is necessary. As~the sky map in
Equation~(\ref{eq:map}) is obtained by summing the sky localization error regions of the BBH detections, the~angular resolution is determined by the diffraction-limited spot
size of the LIGO~detectors:
\begin{equation}
\theta_{\rm res}=\frac{c}{2df}\,,
\label{eq:resolution}
\end{equation}
where $d$ is the typical separation of the detectors in the network, $c$ is the speed of light, and $f$ is the frequency of the measurement. Assuming a typical
frequency of 200 Hz for the detector sensitivity and a LIGO-Virgo detector distance $d\sim 7000$ km, a crude estimate of the minimum map angular resolution
is $\theta_{\rm res}\sim 3^\circ$, or $\sigma\sim 1/30$, implying a high-$l$ cut-off of $l_{\rm max}\sim 30$. 

In the following analysis, for the sake of simplicity, we assume the probability weights in Equation~(\ref{eq:map}) to be constant, i.e., we assume that the
sensitivity of the LIGO-Virgo detector network does not depend on the sky position (see Ref.~\cite{Stiskalek:2020wbj} for a more refined analysis and a discussion
on the effects of detector sensitivity on isotropic test of GW detections). An additional, possible modification of Equation~(\ref{eq:CF}) is due to the different
sensitivities of the GW detector network across the O1-O2 epochs and the varying number of detectors observing each BBH event in the sample. These systematics can
be eliminated, at least partially, by comparing the observed CF $C_{\rm obs}(\theta)$ to a CF, which is computed from a set of $\mathcal{N}$ reference maps $M_{{\rm
ref},k}(\chi,\varphi)$, $k=1,\dots\mathcal{N}$, obtained by uniformly distributing the observed BBH sky localization error regions in the sky. A more refined
analysis could be performed by injecting a population of simulated BBH signals with a uniform angular distribution and then creating the reference map by recovering
the sky localization error regions of these injections with the GW network in the same configuration as in the real case. While this procedure would produce a more
rigorous CF estimate than the one considered here, we consider it beyond the scope of this paper due to the small sample of BBH detections and the illustrative
purpose of our analysis. We plan to revisit this procedure in a future work. 

\section{Results}\label{sec:Results}

\begin{figure}[htbp]
 \centering
 \includegraphics[width=120mm]{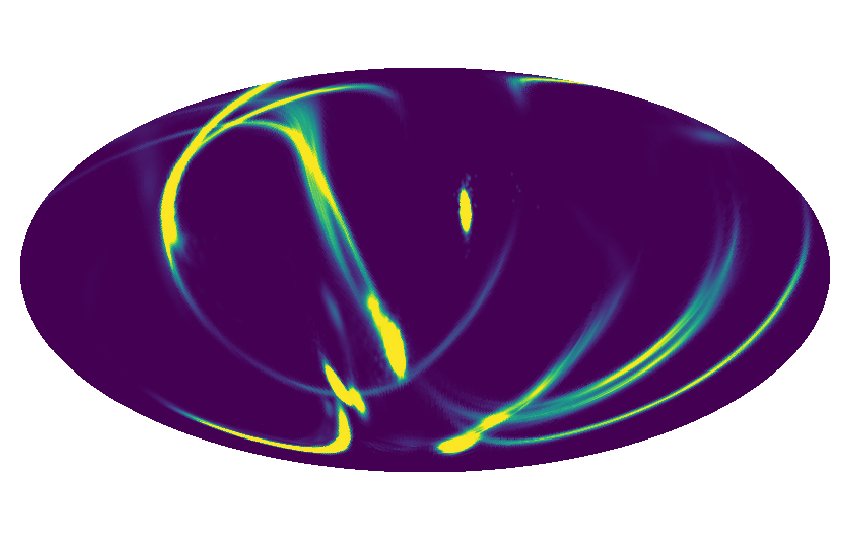}
 \caption{Heat sky map of the combined O1-O2 LIGO-Virgo detections, $M_{\rm obs}(\chi,\varphi)$. The color scale denotes the probability density of sky localization (yellow to blue: high to low, normalized to $\sum_j M(p_j)=0.1$ for better visualization purposes).}
 \label{fig:map}
\end{figure}

We used the public sky localizations of the O1-O2 LIGO-Virgo BBH detections from the GW Open Science Center \cite{open_science} and the open source Healpy package
\cite{healpy} to compute the CF. The sky localization error regions of the BBH detections came with different resolutions. We first rescaled each map to an {\tt
NSIDE}  resolution of 256, corresponding to a pixel angular resolution of $\theta_{\rm pix}\sim 0.23^{\circ}\ll\theta_{\rm res}$. Choosing~different values of the
map resolution affected the final results only by a few percent. We then created the map $M_{\rm obs}(\chi,\varphi)$ in Equation~(\ref{eq:map}) by summing the sky
localization error regions of each BBH event and normalizing to the number of detections, such as $\sum_j M_{\rm obs}(p_j)=1$, where $p_j$ denotes the
$j^{\text{th}}$ pixel. A Mollweide representation of $M_{\rm obs}(\chi,\varphi)$ in Equation~(\ref{eq:map}) is shown in Figure~\ref{fig:map}. 

The $M_{\rm obs}(\chi,\varphi)$ map was treated as a heat map, and the Healpy function {\tt map2alm} was used to compute $a_{lm}$. The coefficients of the Legendre
expansion in Equation~(\ref{eq:CF}) were then obtained by summing the $|a_{lm}|^2$ in $m$. We followed the same procedure to compute the CF from reference maps
$M_{{\rm ref},k}(\chi,\varphi)$ used to test possible angular correlation signatures in the CF. 

\begin{figure}[htbp]
 \centering
 \includegraphics[width=120mm]{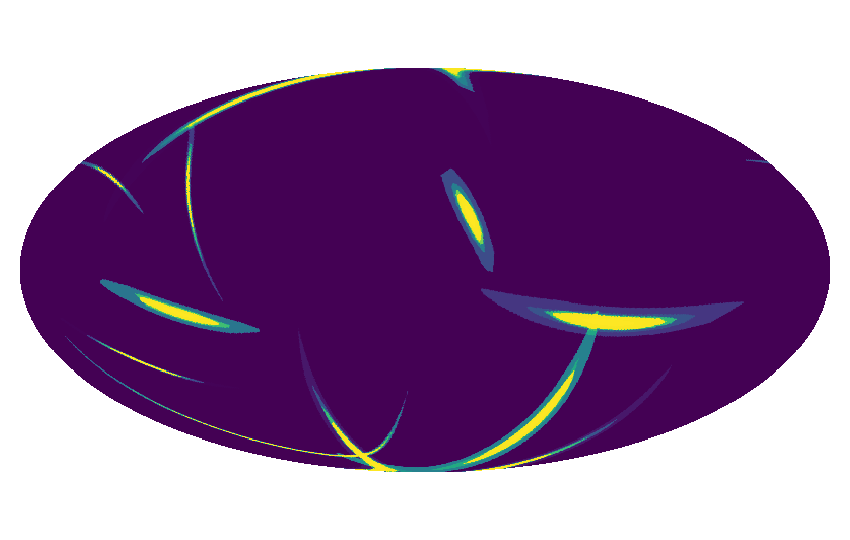}
 \caption{Example of a synthetic sky map for Model B. The (arbitrary) color scale denotes the probability density of sky localization (yellow to blue: high to low, normalized to $\sum_j M(p_j)=0.1$ for better visualization purposes)}
 \label{fig:map_synt}
\end{figure}

In our analysis, we compared the observed CF $C_{\rm obs}(\theta)$ to two reference CFs. The first CF (Model A), $C_{{\rm ref},A}(\theta)$, was obtained by
averaging the CF of 500 artificial maps, each obtained by randomly rotating the maps of each single BBH detection in the sky by arbitrary $\chi$ and $\varphi$
angles. The second CF (Model B), $C_{{\rm ref},B}(\theta)$, was obtained by averaging 500 synthetic maps, each consisting of 10 elliptically-shaped sky localization
error regions with random orientation and uniformly distributed in the sky. The~sky localization areas of these artificial events were chosen such that their
semi-axes were $R\cdot(x,1/x)$, where~$x$ is uniformly distributed in $(0,10)$, and their area $\pi R^2$ was drawn from a lognormal distribution with mean (standard
deviation) equal to the mean (standard deviation) of the sky localization areas of the observed events. Probability distribution contours of each of these
artificial sky localization areas were simulated by superimposing 100 regions built as described above and radius decreasing as $f_n(R)=R\ln(2)/\ln(2+n)$, where
$n=0\dots 99$. The above parameters reproduced sky localization error regions qualitatively similar to the observed BBH error regions while their variations did not
significantly affect the final CF. Both reference maps were normalized to the number of detections in the sample, following the same procedure used for the observed
map. One example of a synthetic map is shown in Figure~\ref{fig:map_synt}. 

\begin{figure}[htbp]
\centering
 \includegraphics[width=125mm]{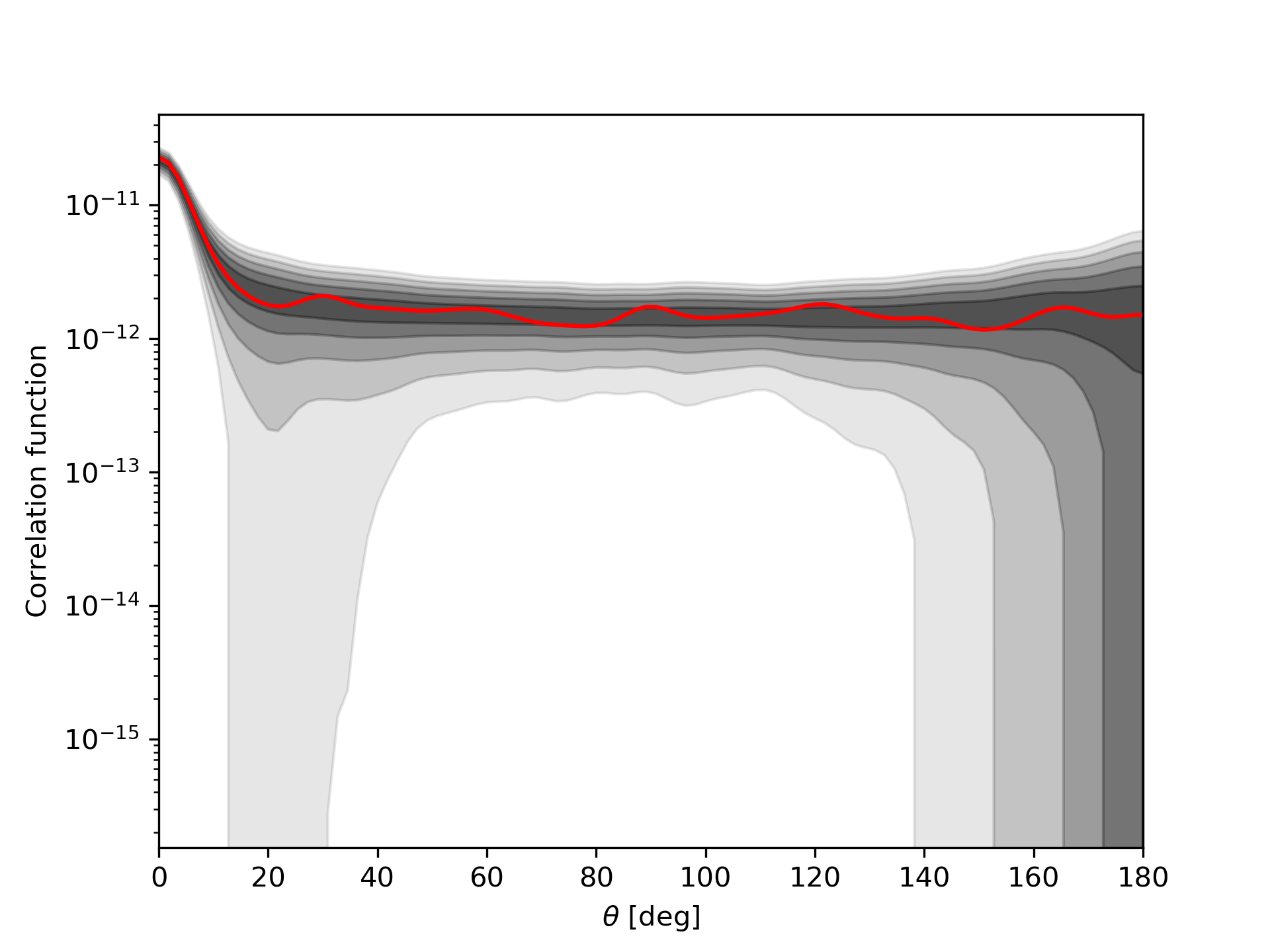}\\ 
 \includegraphics[width=125mm]{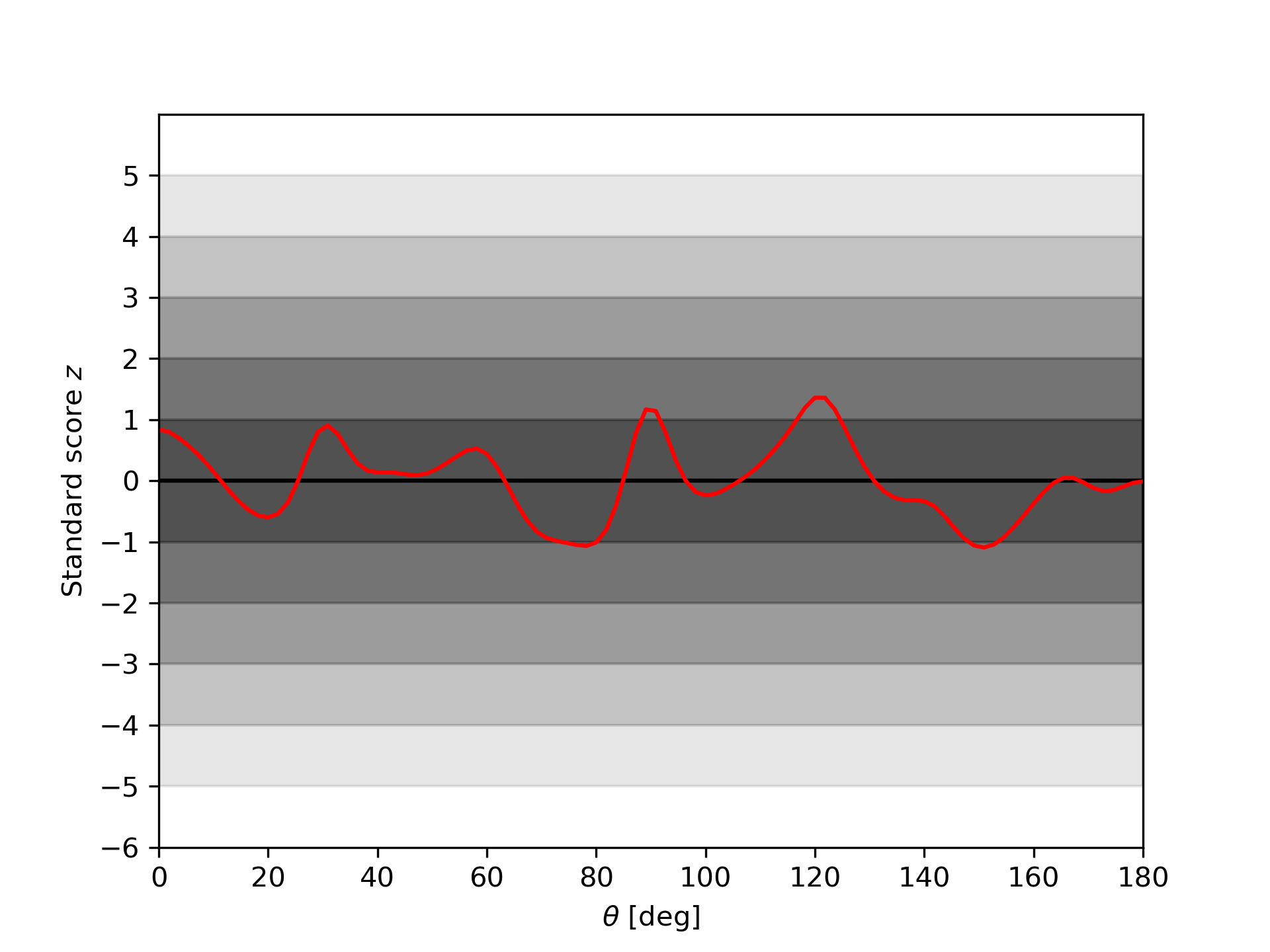}
 \caption{Top: Comparison of the measured CF (red continuous curve) and the reference CF for a set of maps obtained by randomly distributing the LIGO-Virgo observations in the sky (Model A). The~grey-shaded bands denote one- through five-$\sigma$ deviations from the reference CF obtained by averaging over 500 ``random'' maps. Bottom: The observed CF normalized to the reference CF. The observed CF lies within two-$\sigma$ of $C_{{\rm ref},A}(\theta)$.}
\label{fig:corr_rot}
\end{figure}

\begin{figure}[htbp]
\centering
 \includegraphics[width=125mm]{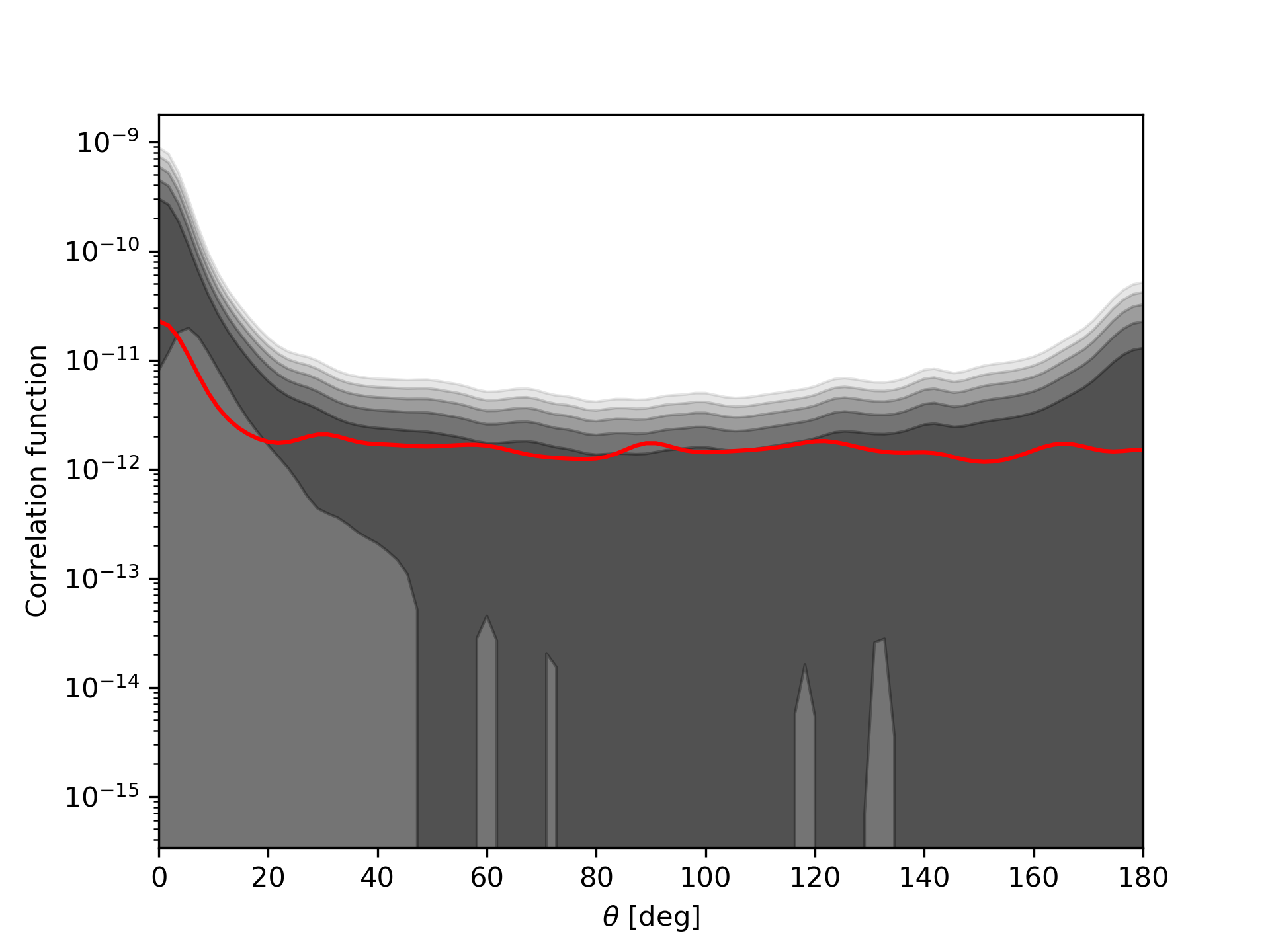}\\
 \includegraphics[width=125mm]{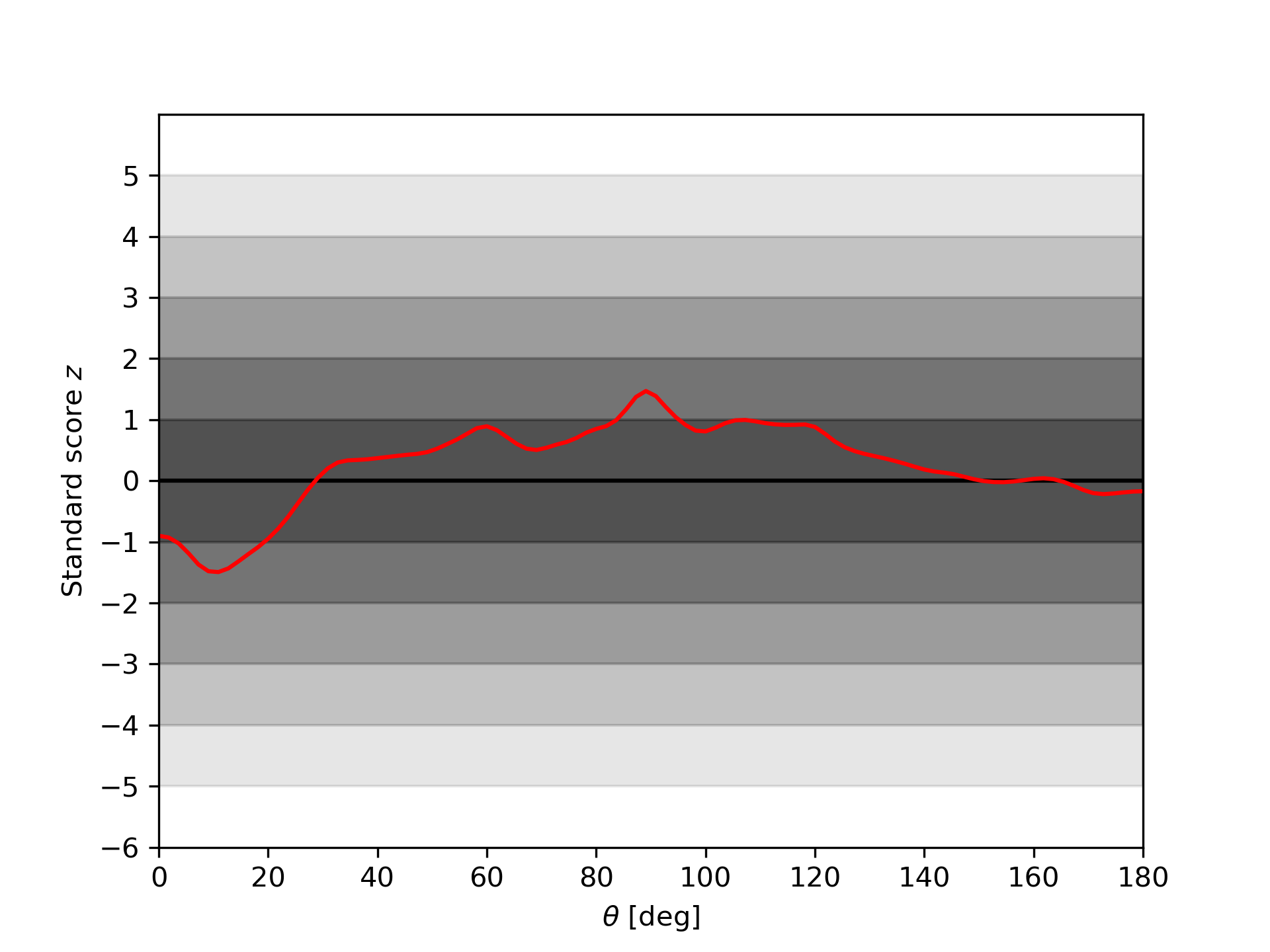}
 \caption{Top: Comparison of the measured CF (red continuous curve) and the CF for the set of synthetic maps of Model B. The~grey-shaded bands denote one- through five-$\sigma$ deviations from the CF averaged over 500 synthetic maps. Bottom: Observed $z$-score of the observed CF for Model B. The~observed CF lies within two-$\sigma$ of $C_{{\rm ref},B}(\theta)$. The deviation at low angular distances is likely due to the approximation used to simulate the synthetic maps.}
\label{fig:corr_synt}
\end{figure}

Figure \ref{fig:corr_rot} shows the observed CF compared to the reference CF for Model A. $C_{\rm obs}(\theta)$ is represented in the top plot by the red curve. The
five grey-shaded bands denote one through five standard deviations from the average of the CF computed on the reference maps $M_{{\rm ref},k}(\chi,\varphi)$ from
Model A, $C_{{\rm ref},A}(\theta)$. The bottom plot shows $C_{\rm obs}(\theta)$ normalized to $C_{{\rm ref},A}(\theta)$. The~observed CF lies entirely in the
two-$\sigma$ band of the reference map, thus showing no excess correlation at any angular scale with respect to a uniform sky distribution of the O1-O2 detections.
No excess correlation at any angular scale could be found when comparing $C_{\rm obs}(\theta)$ to the reference CF for Model B, $C_{{\rm ref},B}(\theta)$. Results
for this model are shown in Figure~\ref{fig:corr_synt}. The observed CF lied entirely within the 2-$\sigma$ error band of $C_{{\rm ref},B}(\theta)$. At small
angular scales, $C_{\rm obs}(\theta)$ showed a lack of correlation compared to $C_{{\rm ref},B}(\theta)$. This mismatch was likely due to the crude approximation
used to simulate the artificial maps. As can be seen from Figure~\ref{fig:map}, the~LIGO-Virgo sky localization error regions of observed BBH events are not perfect
ellipses. Even if they were, their~ellipticity would not follow a uniform distribution in their semi-axis ratio. Finally, drawing~samples from a lognormal
distribution of sky localization areas did not accurately represent the observed distribution of sky localizations in O1 and O2. A much more accurate estimate of
BBH events' angular correlations could be obtained by simulating realistic sky maps by injecting, recovering,~and localizing events according to the actual
sensitivity of the GW detector network.

The CF in Equation~(\ref{eq:CF}) can be interpreted as a weighted projection of the spatial two point CF $\xi(r)$. At~small scales, the power-law behavior of the
CF is expected to be:
\begin{equation}
C(\theta)=\left(\frac{\theta}{\theta_0}\right)^{1-\gamma}\,,
\label{eq:small_scale}
\end{equation}
where $\theta_0$ is an angular correlation scale and $\gamma$ is the power-law slope of the spatial two point CF:
\begin{equation}
\xi(r)=\left(\frac{r}{r_0}\right)^{-\gamma}\,,
\label{eq:spatial}
\end{equation}
where $r_0$ is the spatial correlation length. The power-law slope of the BBH distribution can be obtained by fitting $C_{\rm obs}(\theta)$ at small angular scales.
A weighted best fit of Equation~(\ref{eq:small_scale}) from $\theta\sim\theta_{\rm res}$ to $\theta\sim 18^\circ$, where departures from the power-law behavior
become evident, gives for the power-law slope $\gamma= 2.24\pm 0.33$ at three-$\sigma$ confidence level, a value consistent with a uniform distribution of objects,
$\xi(r)\sim r^{-2}$. As a comparison, the power-law slope from Sloan Digital Sky Survey (SDSS) data is $\gamma\sim 1.8$ over the range $0.005^\circ$--$10^\circ$
\cite{sloan}. The VIMOS  Public Extragalactic Redshift Survey (VIPERS) reports $\gamma \sim 1.7-1.8$ for a broad range of galaxy luminosities and stellar masses in
the redshift range $0.5 < z < 1.1$ \cite{vipers}. The VIMOS-VLT  Deep Survey observes a significant redshift evolution of the luminosity dependence of the power-law
slope parameter with $\gamma$ steepening from $\gamma\sim 1.7$ at low redshift to $\gamma\sim 2.4$ for $z\sim 0.9$ and galaxies with high intrinsic luminosity
\cite{vimos-vlt}. In contrast to the SDSS, VIPERS, and VIMOS-VLT results, which point to galaxy clustering in the redshift range of BBH detections, $z\lesssim 0.5$,
our result showed no evidence of clustering at these distance scales. It would be interesting to test whether any evidence of clustering would appear in the data
with more BBH detections and better sample statistic.

\section{Conclusions}\label{sec:Conclusions}

In this short article, we computed the two-dimensional CF of BBH observations in the first and second observation runs of advanced LIGO and Virgo. The CF is
commonly used in large-scale structure astrophysics and precision cosmology to quantify the spatial distribution of an object class population. Similarly, we used
the two-dimensional CF to measure the statistical properties of the BBH coalescence spatial distribution. By comparing the CF of the LIGO-Virgo detections to a
simulated CF from a synthetic sample of sky localizations and a CF obtained by randomly re-orienting the BBH detections, we showed that the distribution of O1-O2
BBH events in the sky was in agreement with a uniform distribution of sources, as previously reported in Ref.~\cite{Stiskalek:2020wbj}. The power-law slope of the
CF was found to be $\gamma= 2.24\pm 0.33$, a value consistent with the upper bound of the power-law slope from galaxy surveys at low redshift $z$.

While the limited number of O1-O2 detections with large sky localization error regions did not allow us to draw any significant physical conclusions, our work lays
the formalism for computing the CF of a class of GW detections. Our analysis was clearly rudimentary and could be improved in many ways. The extension to the tens
of LIGO-Virgo detections in O3 is straightforward. A better estimate for the two-dimensional spatial distribution of BBH coalescence events could be obtained by
comparing the detected CF to a synthetic CF from a realistic population of events as done in Ref.~\cite{Stiskalek:2020wbj}. This could be done by testing the
detected CF against a CF from injection sets consistent with the observed BBH coalescence population and detector network sensitivity. The existence of angular
correlations in the spatial distribution of BBH coalescences could be tested by building CFs for events distributed isotropically in the sky, or at given angular
scales. Comparisons of the observed BBH CF to CFs of anisotropic models for the astrophysical GW background \cite{Jenkins:2018} and other astrophysical objects
could be used to test correlations of BBH events with the spatial distribution of these objects \cite{Banagiri:2020kqd}, test BBH population paradigms, and probe
fundamental physics \cite{Mukherjee:2019wfw,Mukherjee:2018ebj}. Our method could also be extended to include information about the distances of the BBH sources by
computing the three-dimensional CF~\cite{Vijaykumar:2020pzn}. The latter could be compared to CFs obtained from given models of population synthesis, as well as
three-dimensional CFs of other astrophysical objects. With the anticipated higher rate of detections and more accurate sky localizations in future LIGO-Virgo
observing runs, the CF of BBH and other GW-bright sources may prove itself as another useful tool for GW astronomy investigations.

\section{Acknowledgements}\label{sec:Acknowledgements}
This work started as a high-school science project of A.M.\ at Caddo Parish Magnet High School in Shreveport, LA, under the mentorship of M.C. It was first
presented at the LA state science fair in fall 2019. For this project, A.M. was awarded the Association for Women Geoscientists award, the U.S.\ Agency for
International Development (USAID) award, the Office of Naval Research/U.S. Navy/U.S. Marine Corps award, and the NASA Earth System Science Award. A.M. thanks the
teachers and mentors at the Caddo Parish Magnet High School for their continuous support. M.C.\ is supported by the U.S.\ National Science Foundation grant
PHY-1921006. LIGO was constructed and is operated by the California Institute of Technology and Massachusetts Institute of Technology with funding from the U.S.\
National Science Foundation under grant PHY-0757058. Virgo~is funded by the French Centre National de la Recherche Scientifique (CNRS), the Italian Istituto
Nazionale di Fisica Nucleare (INFN), and the Dutch Nikhef, with contributions by Polish and Hungarian institutes. The~authors would like to thank their LIGO
Scientific Collaboration and Virgo Collaboration colleagues for their help and useful comments, in particular Vuk Mandic, Kentaro Mogushi, Ryan Quitzow-James, Joe
Romano, and Mairi Sakellariadou. The authors are grateful for computational resources provided by the LIGO Laboratory and supported by the U.S.\ National Science
Foundation grants PHY-0757058 and PHY-0823459, as well as resources from the Gravitational Wave Open Science Center, a service of LIGO Laboratory, the LIGO
Scientific Collaboration, and the Virgo Collaboration. This research made use of data, software, and web tools obtained from the Gravitational Wave Open Science
Center and openly available at \url{https://www.gw-openscience.org}. Healpy is licensed under the GNU General Public License. This manuscript has been assigned LIGO
Document Control Center Number LIGO-P2000157.


\begin{thebibliography}{999}
% Reference 1
\bibitem{TheLIGOScientific:2014jea} 
 Aasi, J.; ~et~al. [LIGO Scientific Collaboration]. 
 Advanced LIGO.
 \emph{Class. Quant. Grav.} \textbf{2015}, \emph{32}, 074001. 
 doi:10.1088/0264-9381/32/7/074001.
 %%CITATION = doi:10.1088/0264-9381/32/7/074001;%%

%\cite{Acernese:2015gua}
\bibitem{Acernese:2015gua}
 Acernese, F. [Virgo Collaboration].
 The Advanced Virgo detector.
 \emph{J. Phys. Conf. Ser.} \textbf{2015}, \emph{610}, 012014.
 doi:10.1088/1742-6596/610/1/012014.
 %%CITATION = doi:10.1088/1742-6596/610/1/012014;%%

%\cite{Abbott:2016blz}
\bibitem{Abbott:2016blz} 
 Abbott, B.P.; ~et~al. [LIGO Scientific and Virgo Collaborations].
 Observation of Gravitational Waves from a Binary Black Hole Merger.
 \emph{Phys. Rev. Lett.} \textbf{2016}, \emph{116}, 061102.
 doi:10.1103/PhysRevLett.116.061102.
 %%CITATION = doi:10.1103/PhysRevLett.116.061102;%%
 
%\cite{TheLIGOScientific:2017qsa}
\bibitem{TheLIGOScientific:2017qsa} 
 Abbott, B.P.; ~et~al. [LIGO Scientific and Virgo Collaborations].
 %``GW170817: Observation of Gravitational Waves from a Binary Neutron Star Inspiral,''
 \emph{Phys. Rev. Lett.} \textbf{2017}, \emph{119}, 161101.
 doi:10.1103/PhysRevLett.119.161101.
 %%CITATION = doi:10.1103/PhysRevLett.119.161101;%%
 %191 citations counted in INSPIRE as of 01 Dec 2017 
 
%\cite{GBM:2017lvd}
\bibitem{GBM:2017lvd} 
 Abbott, B.P.; ~et~al. [LIGO Scientific and Virgo and Fermi GBM and INTEGRAL and IceCube and IPN and Insight-Hxmt and ANTARES and Swift and Dark Energy Camera GW-EM and DES and DLT40 and GRAWITA and Fermi-LAT and ATCA and ASKAP and OzGrav and DWF (Deeper Wider Faster Program) and AST3 and CAASTRO and VINROUGE and MASTER and J-GEM and GROWTH and JAGWAR and CaltechNRAO and TTU-NRAO and NuSTAR and Pan-STARRS and KU and Nordic Optical Telescope and ePESSTO and GROND and Texas Tech University and TOROS and BOOTES and MWA and CALET and IKI-GW Follow-up and H.E.S.S. and LOFAR and LWA and HAWC and Pierre Auger and ALMA and Pi of Sky and DFN and ATLAS Telescopes and High Time Resolution Universe Survey and RIMAS and RATIR and SKA South Africa/MeerKAT Collaborations and AstroSat Cadmium Zinc Telluride Imager Team and AGILE Team and 1M2H Team and Las Cumbres Observatory Group and MAXI Team and TZAC Consortium and SALT Group and Euro VLBI Team and Chandra Team at McGill University].
 Multi-messenger Observations of a Binary Neutron Star Merger.
 \emph{Astrophys. J.} {\bf 2017}, \emph{848}, L12.
 doi:10.3847/2041-8213/aa91c9.
 %%CITATION = doi:10.3847/2041-8213/aa91c9;%%
 %1062 citations counted in INSPIRE as of 10 Nov 2019
 
%\cite{gracedb}
\bibitem{gracedb} 
 GraceDB--The Gravitational-wave Candidate Event Database. Available online: 
 \url{http://gracedb.ligo.org}. 

%\cite{Akutsu:2018axf}
\bibitem{Akutsu:2018axf} 
 Akutsu, T.; ~et~al. [KAGRA Collaboration].
 KAGRA: 2.5 Generation Interferometric Gravitational Wave Detector.
 \emph{Nat. Astron.} {\bf 2019}, \emph{3}, 35. 
 doi:10.1038/s41550-018-0658-y.
 %%CITATION = doi:10.1038/s41550-018-0658-y;%%
 %32 citations counted in INSPIRE as of 10 Nov 2019
 
%\cite{LIGO-India}
\bibitem{LIGO-India}
 The Laser Interferometer Gravitational-Wave Observatory (LIGO)-India. Available online: \url{http://www.ligo-india.in}. 

%\cite{LISA}
\bibitem{LISA}
 LISA: Laser Interferometer Space Antenna.~A Proposal in Response to the ESA Call for L3 Mission Concepts. Available online: \url{https://www.elisascience.org/files/publications/LISA\_L3\_20170120.pdf}.


%\cite{Hobbs:2009yy}
\bibitem{Hobbs:2009yy} 
 Hobbs, G.; Archibald, A.; Arzoumanian, Z.; Backer, D.; Bailes, M.; Bhat, N.D.R.; Burgay, M.; \mbox{Burke-Spolaor, S.;} Champion, D.; Cognard, I.; et al. 
 The international pulsar timing array project: using pulsars as a gravitational wave detector.
 \emph{Class. Quant. Grav.} {\bf 2010}, \emph{27}, 084013. 
 doi:10.1088/0264-9381/27/8/084013.
 %%CITATION = doi:10.1088/0264-9381/27/8/084013;%%
 %276 citations counted in INSPIRE as of 16 Nov 2019

%\cite{LIGOScientific:2018mvr}
\bibitem{LIGOScientific:2018mvr} 
 Abbott, B.P.; ~et~al. [LIGO Scientific and Virgo Collaborations].
 GWTC-1: A Gravitational-Wave Transient Catalog of Compact Binary Mergers Observed by LIGO and Virgo during the First and Second Observing Runs. 
 \emph{Phys. Rev. X} {\bf 2019}, \emph{9}, 031040. 
 doi:10.1103/PhysRevX.9.031040.
 %%CITATION = doi:10.1103/PhysRevX.9.031040;%%
 %532 citations counted in INSPIRE as of 10 Nov 2019

%\cite{LIGOScientific:2019fpa}
\bibitem{LIGOScientific:2019fpa} 
 Abbott, B.P.; ~et~al. [LIGO Scientific and Virgo Collaborations].
 Tests of General Relativity with the Binary Black Hole Signals from the LIGO-Virgo Catalog GWTC-1. \emph{arXiv} \textbf{2019}, 
 arXiv1903.04467.
 %%CITATION = ARXIV:1903.04467;%%
 %64 citations counted in INSPIRE as of 10 Nov 2019

%\cite{LIGOScientific:2018jsj}
\bibitem{LIGOScientific:2018jsj} 
 Abbott, B.P.; ~et~al. [LIGO Scientific and Virgo Collaborations].
 Binary Black Hole Population Properties Inferred from the First and Second Observing Runs of Advanced LIGO and Advanced Virgo.
 \emph{Astrophys. J.} {\bf 2019}, \emph{882}, L24.
 doi:10.3847/2041-8213/ab3800.
 %%CITATION = doi:10.3847/2041-8213/ab3800;%%
 %164 citations counted in INSPIRE as of 10 Nov 2019

%\cite{LIGOScientific:2019eut}
\bibitem{LIGOScientific:2019eut} 
 Abbott, B.P.; ~et~al. [LIGO Scientific and Virgo Collaborations].
 Model comparison from LIGO-Virgo data on GW170817's binary components and consequences for the merger remnant. \emph{arXiv} \textbf{2019}, 
 arXiv:1908.01012.
 %%CITATION = ARXIV:1908.01012;%%
 %7 citations counted in INSPIRE as of 10 Nov 2019

%\cite{Abbott:2019yzh}
\bibitem{Abbott:2019yzh} 
 Abbott, B.P.; ~et~al. [LIGO Scientific and Virgo Collaborations].
 A gravitational-wave measurement of the Hubble constant following the second observing run of Advanced LIGO and Virgo. \emph{arXiv} \textbf{2019}, 
 arXiv:1908.06060.
 %%CITATION = ARXIV:1908.06060;%%
 %7 citations counted in INSPIRE as of 10 Nov 2019

%\cite{Abbott:2019prv}
\bibitem{Abbott:2019prv} 
 Abbott, B.P.; ~et~al. [LIGO Scientific and Virgo Collaborations].
 All-sky search for short gravitational-wave bursts in the second Advanced LIGO and Advanced Virgo run.
 \emph{Phys. Rev. D} {\bf 2019}, \emph{100}, 024017.
 doi:10.1103/PhysRevD.100.024017.
 %%CITATION = doi:10.1103/PhysRevD.100.024017;%%
 %5 citations counted in INSPIRE as of 10 Nov 2019

%\cite{Abbott:2019heg}
\bibitem{Abbott:2019heg} 
 Abbott, B.P.; ~et~al. [LIGO Scientific and Virgo Collaborations].
 All-sky search for long-duration gravitational-wave transients in the second Advanced LIGO observing run.
 \emph{Phys. Rev. D} {\bf 2019}, \emph{99}, 104033.
 doi:10.1103/PhysRevD.99.104033.
 %%CITATION = doi:10.1103/PhysRevD.99.104033;%%
 %5 citations counted in INSPIRE as of 10 Nov 2019

%\cite{Pisarski:2019vxw}
\bibitem{Pisarski:2019vxw} 
 Abbott, B.P.; ~et~al. [LIGO Scientific and Virgo Collaborations].
 All-sky search for continuous gravitational waves from isolated neutron stars using Advanced LIGO O2 data.
 \emph{Phys. Rev. D} {\bf 2019}, \emph{100}, 024004.
 doi:10.1103/PhysRevD.100.024004.
 %%CITATION = doi:10.1103/PhysRevD.100.024004;%%
 %21 citations counted in INSPIRE as of 10 Nov 2019

%\cite{Abbott:2019pxc}
\bibitem{Abbott:2019pxc} 
 Abbott, B.P.; ~et~al. [LIGO Scientific and Virgo Collaborations].
 An Optically Targeted Search for Gravitational Waves emitted by Core-Collapse Supernovae during the First and Second Observing Runs of Advanced LIGO and Advanced Virgo. \emph{arXiv} \textbf{2019},
 arXiv:1908.03584.
 %%CITATION = ARXIV:1908.03584;%%
 %1 citations counted in INSPIRE as of 10 Nov 2019

%\cite{Abbott:2019dxx}
\bibitem{Abbott:2019dxx} 
 Abbott, B.P.; ~et~al. [LIGO Scientific and Virgo Collaborations].
 Search for Transient Gravitational-wave Signals Associated with Magnetar Bursts during Advanced LIGO's Second Observing Run.
 \emph{Astrophys. J.} {\bf 2019}, \emph{874}, 163.
 doi:10.3847/1538-4357/ab0e15.
 %%CITATION = doi:10.3847/1538-4357/ab0e15;%%
 %6 citations counted in INSPIRE as of 10 Nov 2019
 
%\cite{White:1994}
\bibitem{White:1994}
 White, M.; Scott, D.; Silk, J. 
 Anisotropies in the Cosmic Microwave Background.
 \emph{Annu. Rev. Astron. Astrophys.} {\bf 1994}, \emph{32}, 319--370.

%\cite{Stiskalek:2020wbj}
\bibitem{Stiskalek:2020wbj}
Stiskalek, R.; Veitch, J.; Messenger, C. Are stellar mass binary black hole mergers isotropically distributed? \emph{arXiv} \textbf{2020},
%``Are stellar mass binary black hole mergers isotropically distributed?,''
arXiv:2003.02919.
%0 citations counted in INSPIRE as of 27 Apr 2020

%\cite{open_science}
\bibitem{open_science}
 Abbott, R.; ~et~al. [LIGO Scientific and Virgo Collaborations].
 Open data from the first and second observing runs of Advanced LIGO and Advanced Virgo. \emph{arXiv} \textbf{2019},
 arXiv:1912.11716.

%\cite{healpy}
\bibitem{healpy}
 Healpy, a Python Package to Handle Pixelated Data on the Sphere. Available online: 
 \url{healpy.readthedocs.io/en/latest.index.html}.


%\cite{sloan}
\bibitem{sloan}
 Wang, Y.; Brunner, R.J.; Dolence, J.C. 
 The SDSS Galaxy Angular Two-Point Correlation Function. 
 \emph{ Mon.~Not.~R.~Astron.~Soc.} ~{\bf 2013}, \emph{432}, 1961--1979. 

%\cite{vipers}
\bibitem{vipers}
 Marulli, F.; Bolzonella, M.; Branchini, E.; Davidzon, I.; De La Torre, S.; Granett, B.R.; Guzzo, L.; Iovino, A.; Moscardini, L.; Pollo, A.; et al.
 The VIMOS Public Extragalactic Redshift Survey (VIPERS)--Luminosity and stellar mass dependence of galaxy clustering at $0.5 < z < 1.1$.
 \emph{Astron.~Astrophys.}~\textbf{2013}, {\em 557}, A17.

%\cite{vimos-vlt}
\bibitem{vimos-vlt}
 Pollo, A.; Guzzo, L.; F{\`e}vre, O.L.; Meneux, B.; Cappi, A.; Franzetti, P.; Iovino, A.; McCracken, H.J.; \mbox{Marinoni, C.;} Zamorani, G.;~et~al.
 The VIMOS-VLT Deep Survey--Luminosity dependence of clustering at $z\sim 1$.
 \emph{Astron. Astrophys.} {\bf 2006}, \emph{451}, 409.
 doi:10.1051/0004-6361:20054705.

%\cite{Jenkins:2018}
\bibitem{Jenkins:2018}
 Jenkins, A.C.; Sakellariadou, M.; Regimbau, T.; Slezak, E.
 Anisotropies in the astrophysical gravitational-wave background: Predictions for the detection of compact binaries by LIGO and Virgo. 
 \emph{Phys. Rev. D} {\bf 2018}, \emph{98}, 063501.

%\cite{Banagiri:2020kqd}
\bibitem{Banagiri:2020kqd}
Banagiri, S.; Mandic, V.; Scarlata, C.; Yang, K.Z. 
Measuring angular N-point correlations of binary black-hole merger gravitational-wave events with hierarchical Bayesian inference. \emph{arXiv} \textbf{2020}, arXiv:2006.00633.
%0 citations counted in INSPIRE as of 22 Jun 2020

%\cite{Mukherjee:2019wfw}
\bibitem{Mukherjee:2019wfw}
Mukherjee, S.; Wandelt, B.D.; Silk, J. 
Multi-messenger tests of gravity with weakly lensed gravitational waves.
\emph{Phys. Rev. D} {\bf 2020}, \emph{101}, 103509.
doi:10.1103/PhysRevD.101.103509.
%6 citations counted in INSPIRE as of 19 May 2020

%\cite{Mukherjee:2018ebj}
\bibitem{Mukherjee:2018ebj}
Mukherjee, S.; Wandelt, B.D. 
Beyond the classical distance-redshift test: cross-correlating redshift-free standard candles and sirens with redshift surveys. \emph{arXiv} \textbf{2018},
arXiv:1808.06615.
%8 citations counted in INSPIRE as of 19 May 2020

%\cite{Vijaykumar:2020pzn}
\bibitem{Vijaykumar:2020pzn}
Vijaykumar, A.; Saketh, M.; Kumar, S.; Ajith, P.; Choudhury, T.R. 
Probing the large scale structure using gravitational-wave observations of binary black holes. \emph{arXiv} \textbf{2020},
arXiv:2005.01111.
\end{thebibliography}
\end{document}